\documentclass[12pt]{article}
\usepackage{graphicx}
\begin{document}
\bibliographystyle{num}
\title{Spectral properties of nuclear matter}
\baselineskip=1. \baselineskip

\author{     P. Bo\.{z}ek\footnote{Electronic
address~:
piotr.bozek@ifj.edu.pl}\\
Institute of Nuclear Physics, PL-31-342 Cracow, Poland}

\date{\today}

\maketitle

\begin{abstract}
We review self-consistent spectral methods for  nuclear matter calculations.
The  in-medium $T-$matrix approach is 
conserving and thermodynamically consistent. It gives  both the global
and the single-particle properties the system.
The $T-$matrix approximation allows to address the pairing phenomenon in cold
nuclear matter. A generalization of  nuclear matter calculations to the
superfluid phase is discussed and numerical results are presented for this
case. The linear response of a correlated system going beyond the Hartree-Fock+
Random-Phase-Approximation (RPA) 
scheme is studied. The polarization is obtained by
solving a consistent Bethe-Salpeter (BS) equation for the coupling of dressed
nucleons to an external field.
 We find that multipair contributions are important for the 
spin(isospin) response
when the interaction is spin(isospin) dependent.

\end{abstract}

\vskip .4cm
 \section{Introduction}
Nuclear matter is an infinite system of strongly interacting fermions.
Low energy nucleon-nucleon interactions in vacuum  are well known
from scattering experiments. The characteristic
 features of the nuclear force are
the presence of a strongly repulsive core at small distances, 
attraction at moderate distances, and the appearance of a tensor component.
 The properties of nucleons in a strongly
interacting medium are modified. This is taken into account by the dressing of
nucleon propagators by a self-consistently calculated self-energy.
Nonzero imaginary part of the self-energy implies a finite lifetime of
quasiparticles and gives them nontrivial spectral properties;
nucleons propagate off the energy shell.
The description of  nuclear matter in the language of such 
dressed nucleons with a broad spectral function has been the subject of
intensive studies in the last years.

\section{In medium $T$ matrix}

The description of  properties and excitations of a many-body system can
be performed in the language of finite-temperature Green's functions. 
The Green's functions' approach is most easily formulated in the imaginary time
formalism \cite{matsubara}, suitable for formal
presentations and perturbative calculations, but 
the real-time formalism  \cite{keldysh} is
better adapted  for  numerical calculations. In medium propagators (Green's
functions) are dressed by a self-energy term. The physical approximation
enters in the choice of a suitable expression for the self-energy used in
the calculation. Analytical properties of the Green's functions are guaranteed
by  the dispersion relation between the imaginary and real parts of the
self-energy and by the use of  the Dyson equation
\begin{equation}
G^{-1}=G_0^{-1}-\Sigma  \ ,
\label{dyson}
\end{equation}
where $G$ and $G_0$ represent 
the dressed and vacuum propagators respectively and $\Sigma $
is the self-energy.

The choice of a specific form of the self-energy should be adapted to
the  physical
system under consideration, but a  general procedure to derive 
nonperturbative approximations in many-body systems has been proposed 
\cite{Baym}. The self-energy is written as a functional derivative of a 
two-particle irreducible generating functional
$\Sigma=\frac{\delta\Phi}{\delta G}$.

\begin{figure}
\centering
\includegraphics*[width=0.5\textwidth]{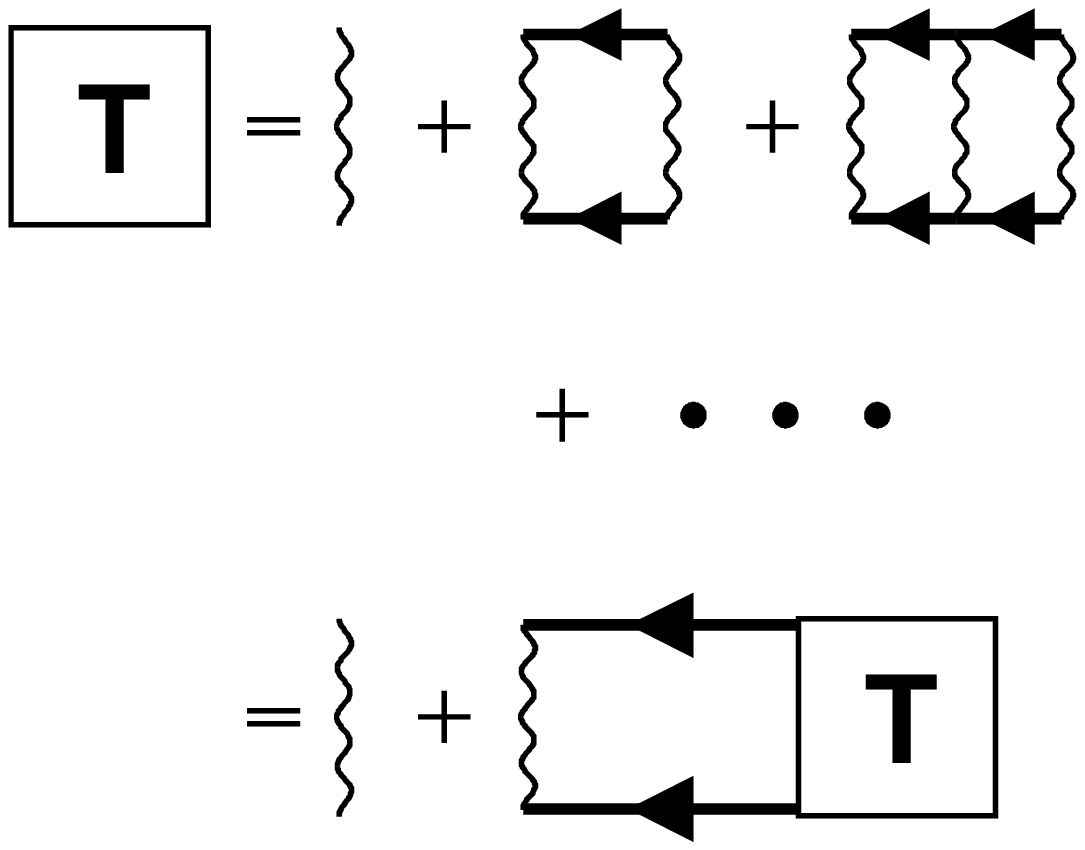}
\caption{Definition of the $T-$matrix as a sum ladder diagrams in the
  interaction.}
\label{tmfig}
\end{figure}

\begin{figure}
\centering
\includegraphics*[width=0.5\textwidth]{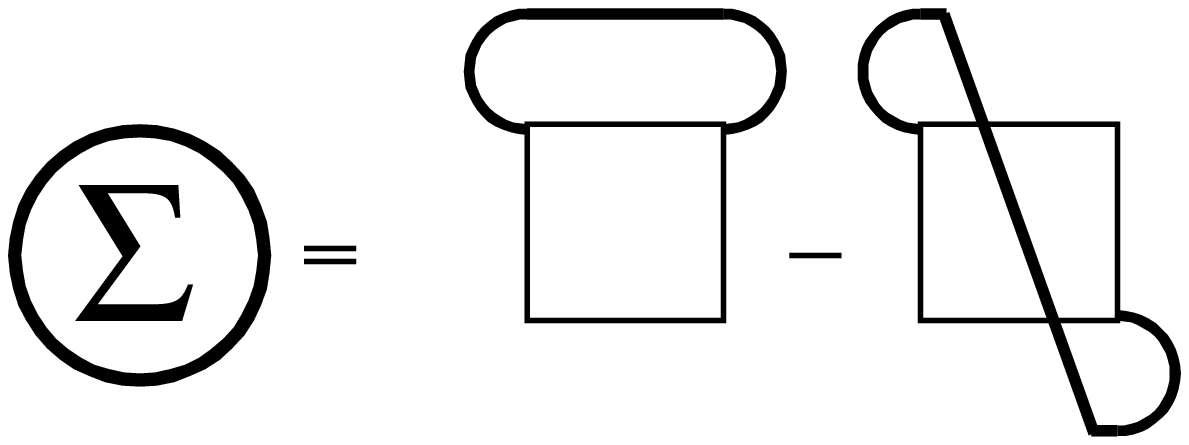}
\caption{The self-energy in the $T-$matrix approximation}
\label{selftm}
\end{figure}

Due to the presence of a short range repulsive core,
ladder diagrams in the free nucleon-nucleon interaction $V$ must be resummed.
The resulting correlated part of the two-particle Green's function is called
the in-medium $T-$matrix \cite{KadanoffBaym} (Fig. \ref{tmfig})
\begin{equation}
T=V+VGGT \ .
\label{tmeq}
\end{equation}
Although the ladder diagrams for the $T$ matrix have the same form  as for the 
Brueckner $G-$matrix the resulting expressions are different. The $T-$matrix
equation is defined for in-medium Green's functions. At zero temperature the
two-nucleon propagator in Fig. \ref{tmfig} represents the propagation a
particle-particle pair (excitations above the Fermi energy) or a hole-hole
pair  (excitations below the Fermi energy), whereas in the Brueckner scheme
the blocking operator forces the two nucleons in the ladder to be of the
particle-particle type. The second, even more important, difference comes 
at the level of the self-energy, which in the $\Phi-$derivable
 $T-$matrix approximation must be taken self-consistently in the form (Fig
\ref{selftm})
\begin{equation}
\Sigma=Tr T G \ .
\label{seq}
\end{equation}
 The self-consistency means that the nucleon propagators at all stages of the
 approximation are dressed by this self-energy. It requires an iterative
 procedure for the solution of the coupled equations (\ref{selftm}),
(\ref{seq}), and (\ref{dyson}). The $T$ matrix
 obtained in a given iteration is used to generated the self-energy and the
 dressed propagators. The dressed propagators are then used to calculate the 
in-medium $T$ matrix in the  next iteration. 
\begin{figure}
\centering
\includegraphics*[width=0.45\textwidth]{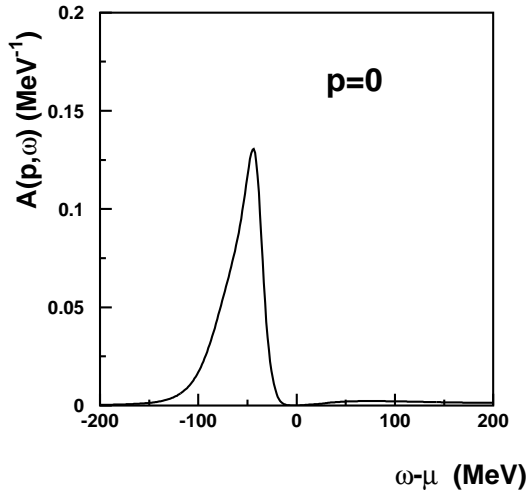}
\caption{The nucleon spectral function obtained in the self-consistent
  $T-$matrix approximation for nuclear matter.}
\label{spectral}
\end{figure}

The numerical calculations are very demanding due to the fact that the
 nucleons are dressed by nontrivial spectral functions coming from the
imaginary part of the self-energy (Fig \ref{spectral}); additional integrations
 over the energy of off-shell nucleons appear.
First results on the self-consistent 
 in-medium $T$ matrix appeared only in the recent years
  \cite{di1,Bozek:1998su}. 
In \cite{di1,Dickhoff:1999yi} the spectral function of the dressed
nucleon was parameterized by three Gaussians and in 
\cite{Dewulf:2000jg,Dewulf:2002gi} using a three poles approximation for the
in-medium propagator.
 The first solution using full spectral
functions for the dressed propagators in nuclear matter was obtained
numerically in ref. \cite{Bozek:1998su} at finite temperature, for a simple
nuclear interaction. This approach, using numerical algorithms significantly
speeding the calculations, was then generalized to realistic interactions
and zero temperature
 \cite{Bozek:2002em,Bozek:2002tz}. The  solution of the
 $T-$matrix scheme at finite temperature 
was also obtained in \cite{Frick:2003sd}.

At low temperatures two  difficulties show up. The first one is
related to the transition to superfluidity in cold nuclear matter. 
This effect can be taken into account quite naturally within suitably 
generalized $T-$matrix approaches. This is the subject of the next section. 
We note however that most of the existing $T-$matrix calculations are done 
for the normal phase of  the nuclear matter also at zero temperature.
The second difficulty with low temperature nuclear matter is technical and is
related to the appearance of  a well defined quasi-particle peak in the
spectral function of the in-medium nucleon. Simple discretization algorithms
for the energy integration break down in that case. The spectral function must
be separated into  background and 
quasi-particle contributions for momenta close to $p_F$. 
At the time of this writing, 
there is only one numerical implementation of this procedure for the 
nuclear matter $T$ matrix (see \cite{Bozek:2002em} for details).

First calculations \cite{di1,Bozek:1998su,Dickhoff:1999yi}  have  shown
that the self-consistency for the self-energies of nucleons is very important.
The effective scattering is reduced in the self-consistent calculation when
compared to an approximation neglecting the imaginary part of the self-energy
\cite{Alm:1996ps}. Also the value of the critical temperature for the
superfluid phase transition goes down when using fully dressed propagators in
the Thouless  criterion for superfluidity.

Further developments have addressed the binding energy in the $T-$matrix 
approximation \cite{Bozek:2001tz,Bozek:2002mw,Dewulf:2002gi,Dewulf:2003nj}.
The role of the correlations, high momentum states, and low energy tails
in the spectral function on binding has been  discussed \cite{Bozek:2002mw}.
Formally, 
one should not expect better results for the binding energy in the $T-$matrix
approximation than in the standard $G-$matrix approach with higher order terms
in the hole line-expansion \cite{3holes}.
 However,  the authors of ref.  \cite{Dewulf:2003nj} claim
that the $T-$matrix calculation, which does not take into account 
ring diagrams contribution, is better adapted to describe nuclear matter in
finite nuclei. An important observation is made  in
\begin{figure}
\centering
\includegraphics*[width=0.4\textwidth]{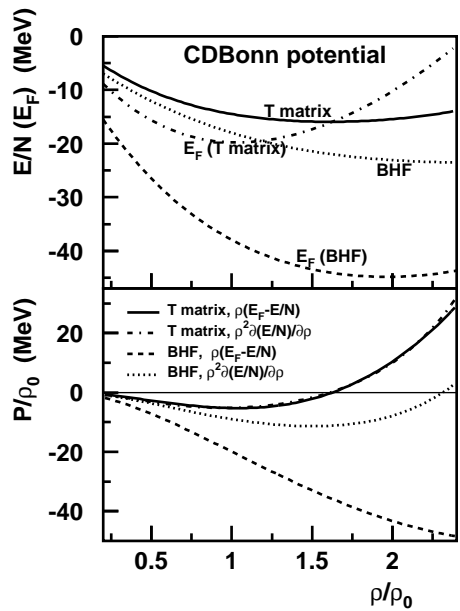}
\caption{The binding energy and the Fermi energy (upper panel) and
two (formally equivalent)
expressions for the pressure (lower panel) as function of density
in the $T-$matrix and the Brueckner-Hartree-Fock calculations.}
\label{press}
\end{figure}
ref. \cite{Bozek:2001tz}, where the role of the thermodynamical consistency of
the $T-$matrix approximation is discussed. As expected from the
$\Phi-$derivability of the scheme, the $T-$matrix calculation gives
consistent results for the  global quantities in the system and its single
particle properties (Fig. \ref{press}). In particular the $T-$matrix results 
 fulfill the  Hugenholz-Van Hove theorem and the Luttinger relation for the
 value of the Fermi momentum.   
In practice, 
such a consistency of 
the single particle energy can be obtained within the Brueckner
theory only approximately by taking  rearrangement terms to a given order.

The single-particle properties obtained in the self-consistent iteration of
the $T-$matrix scheme represent a reliable estimate of the optical potential
and of the single-particle width in medium \cite{Bozek:2002em}. 
Nuclear matter in the non-superfluid phase behaves as a Fermi liquid, with
the standard phase-space scaling of the scattering width with the distance to 
the Fermi energy and with the temperature \cite{Bozek:2002em}.
 The spectral function in finite
nuclei has been estimated taking into account the local density and the
neutron-proton asymmetry \cite{Bozek:2003wh} and compared to experimental data.
Thermodynamic properties of the asymmetric \cite{Frick:2004th} and pure
neutron nuclear matter \cite{Bozek:2002ry}
 have been discussed. We note that the $T-$matrix calculation can
be performed very easily 
at finite temperature \cite{Bozek:1998su,Bozek:2002em,Frick:2003sd} 
(it is even simpler than at zero temperature).

\section{Pairing}

\begin{figure}[hb]
\centering
\includegraphics*[width=0.45\textwidth]{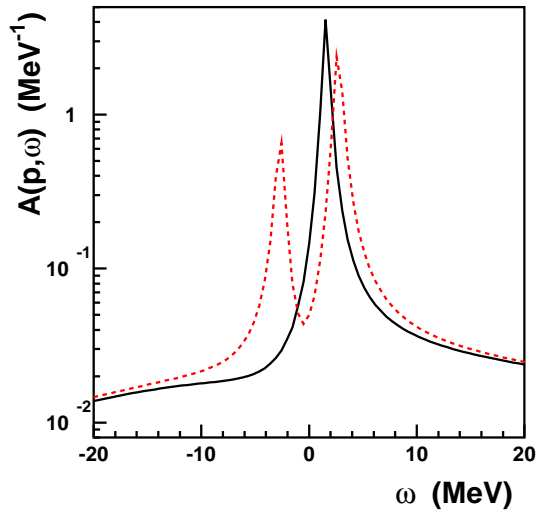}
\caption{The spectral function in the superfluid phase, including the the
  diagonal self-energy (solid line) and both the diagonal and anomalous
  self-energies (dashed line).}
\label{specsf}
\end{figure}
 
A general property of fermions interacting through 
an attractive potential is the
transition to a superfluid state at low temperatures \cite{BCS}.
 This phenomenon is
usually not taken into account in nuclear matter studies, because the value of
the superfluid gap, as extracted from the properties of finite nuclei,
  is expected to be small.  On the other hand naive estimates of the
  neutron-proton pairing gap yield a value of several MeVs 
\cite{Vonderfecht:1991}. This value, obtained from a mean-field gap equation,
 is
modified by many-body effects.

A fundamental approach in the study of superfluid nuclear matter should aim at 
the description of the  cold nuclear matter system dealing with a strong 
repulsive core in the interaction  and at the same time with 
 the formation of the superfluid state.
A generalization of nuclear matter calculations must 
include the usual ladder diagram resummation in the normal phase. At the same
time it should describe the formation of the superfluid long-range order in the
two-particle correlations. Such correlations appear as a singularity of the
$T$ matrix for $T=T_c$ at the energy of twice the fermionic chemical
potential. This corresponds to the formation of a two-fermion bound state. 
At temperatures below $T_c$ such fermionic (Cooper)
 pairs have a binding energy 
allowing for the condensation and for the  formation of a superfluid order
 parameter.

\begin{figure}[hb]
\centering
\includegraphics*[width=0.45\textwidth]{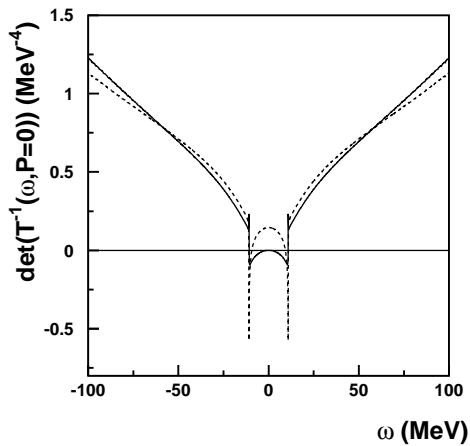}
\caption{  Inverse of the $T-$matrix in the pairing (deuteron) channel in the
  superfluid phase. The usual $T$ matrix (dashed line) and the generalized
  $T$ matrix with   off-diagonal components  (solid line) are shown
 \cite{Bozek:2001nx}.
The generalized $T$ matrix has a singularity 
at the Fermi energy  for zero total momentum of the
  pair.}
\label{sfoso}
\end{figure}
 
In ref. \cite{Bozek:1999rv} a first attempt to generalize  nuclear  
matter calculation to the superfluid phase has been discussed. At small
temperatures a superfluid state forms leading to nonzero expectations of the
off-diagonal (anomalous) Green's functions and self-energies 
\cite{schrieffer}. The off-diagonal self-energy is  obtained from a gap 
equation but with dressed propagators. The $T-$matrix in the superfluid must
be  constructed in a way  to conserve the singularity at twice the
chemical potential also for $T<T_c$. Such a scheme was considered in 
\cite{Bozek:1999rv}, giving the first calculation of the superfluid nuclear
matter problem, including both the ladder 
resummation of the nuclear  interaction and 
  the superfluid properties obtained with a gap equation modified
 by many-body effects. The nucleon propagator in the superfluid is dressed
by a diagonal self-energy $\Sigma$ coming from a modified $T-$matrix
approximation and an off-diagonal anomalous self-energy $\Delta$; it has
two poles on both sides  the Fermi energy (Fig. \ref{specsf}).
 The most interesting result of
this first, and up to  now unique, study is the observation of 
a strong reduction of the superfluid gap in the correlated nuclear matter
as compared to results from a mean-field gap equation  \cite{Bozek:1999rv}.
This effect was analyzed in details \cite{Bozek:2000fn} and can be explained
by an effective modification of the density of states at the Fermi energy.
The reduction of the superfluid energy gap is especially important for
neutron-proton pairing close to the gap closure in symmetric nuclear matter
\cite{Bozek:2002jw}.

Possible  generalizations of the $T-$matrix approximation to the  
superfluid phase have been discussed by Haussmann \cite{Haussman2}.
The simplest scheme, which is $\Phi-$derivable and includes the
required ingredients,
 i.e. the ladder diagrams resummation in the diagonal self-energy and
the gap equation for the off-diagonal self-energy, requires
the introduction of an additional $T-$matrix describing  two-particle
correlations  for anomalous propagators.  The generalized $T-$ matrix has 
a singularity at twice the Fermi momentum for $T\le T_c$ (Fig. \ref{sfoso})
 and the 
imaginary part of
the self-energy has a gap around the Fermi energy, corresponding to an energy
gap for possible excitations in a scattering \cite{Bozek:2001nx}. This 
last, important  property of the approximation does not hold for the simple
scheme discussed in \cite{Bozek:1999rv}. Explicit calculations show that
corrections to the gap equation coming from ladder diagrams in the
off-diagonal self-energy are small 
\cite{Bozek:2001nx}. These small corrections make the 
superfluid gap energy dependent.

\section{Linear response with dressed vertices}

Processes occurring in the  dense nuclear matter are modified 
by medium effects \cite{Knoll:1996nz}. This happens  for neutrino rates
in neutron stars  and for 
particle or photon  emission in hot nuclear matter.
 In particular, the effect
of the off-shell propagation of nucleons in the medium is
important for the subthreshold particle production in heavy ion
collisions. The role of  correlations for the neutrino emission could
be especially
important for processes in hot stars.

If the interaction with an external perturbation
is small the 
dynamics of the system in the external field
 can be described in terms of linear response
functions. The response functions incorporate all the correlations due
to the self-interaction of particles in the system.
For normal Fermi liquids the response function to long-wavelength
perturbations
 can be calculated within the Fermi liquid theory \cite{pines}.
 However, if one is
 interested in nuclear systems at higher temperatures or for
 perturbations with large momentum  a different approach is required.
The Fermi liquid theory does not allow for multipair excitations 
of the system in the external potential. Such multipair excitations 
are important for higher energies and momenta of the external
perturbation
or in the presence of tensor interactions \cite{olpe}.

\begin{figure}
\centering
\includegraphics*[width=0.5\textwidth]{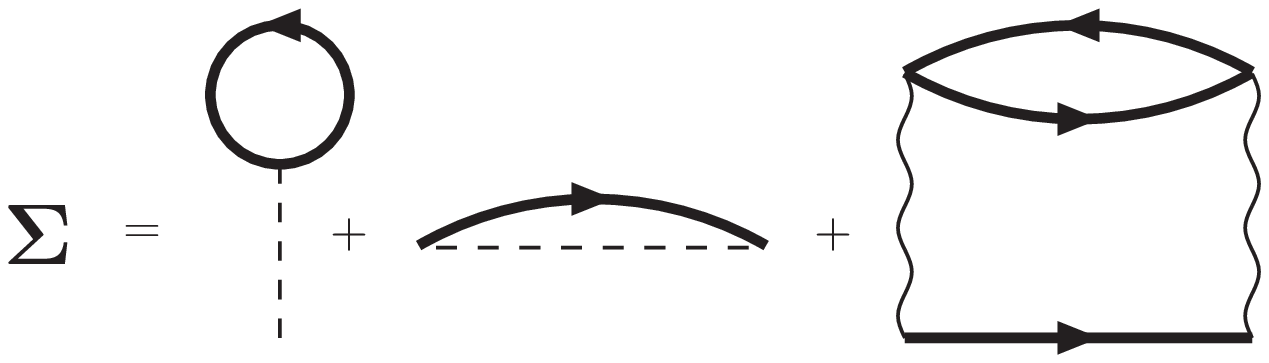}
\caption{Diagrams for the self-energy. The first two diagrams are the
  Hartree-Fock contribution for the Gogny interaction.
 The last diagram is the contribution of the residual
  interaction in  the second order.} 
\label{selffig}
\end{figure}

We take the 
 interactions between the nucleons as a sum of a  
 mean-field interaction that 
 is based on the Gogny parameterization \cite{gogny}
 and a residual interaction (scalar or isospin dependent). 
The isospin dependent residual 
interaction is obtained from the scalar one  multiplying by the factor 
$\frac{1}{2}(1+\tau_1\tau_2)$.
The self-energy is taken in the second direct Born
approximation for the residual interaction (Fig. \ref{selffig}).
The residual interaction induces a finite width to  nucleon
excitations in the medium. Such a dressing of nucleons is expected in
any approach going beyond the simple mean-field, e.g. the $T-$matrix 
approximation discussed previously.

When the description of the correlated system goes beyond the
mean-field approximation  the difficulty involved in  a 
consistent calculation of the response
function is severely increased \cite{bonitz,faleev}. 
A naive calculation of the 
 polarization bubble using dressed propagators
\begin{equation}
\label{polsimp}
\Pi=Tr\Gamma_0 G_{ph} \Gamma_0 \ , 
\end{equation}
where $G_{ph}$ is the particle-hole propagator with dressed 
nucleons and $\Gamma_0$ is the free vertex for the coupling of the nucleon to
an external field,
is a very bad estimate for the response function \cite{pbpl}. In
 particular, it
severely violates the $\omega$-sum rule  \cite{henn}.

\begin{figure}
\centering
\includegraphics*[width=0.5\textwidth]{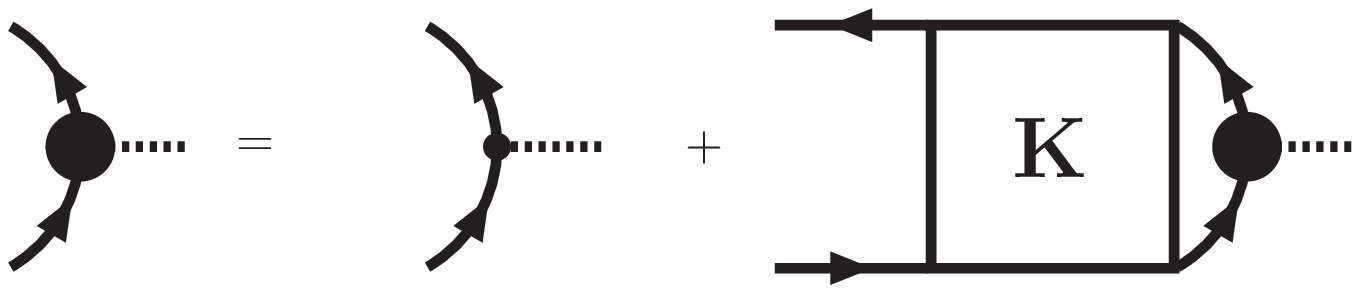}
\caption{The Bethe-Salpeter equation for the dressed vertex. The
  particle-hole
irreducible kernel $K$ is denoted by the box and the fat and the
  small dots denote  respectively
  the dressed and the bare vertices for the coupling of the external
  field to the nucleon. 
}
\label{bsfig}
\end{figure}

\begin{figure}
\centering
\includegraphics*[width=0.3\textwidth]{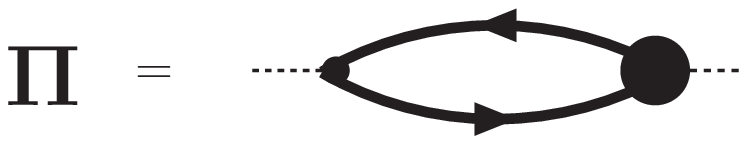}
\caption{The polarization function expressed using the dressed vertex
  for the coupling of the external field to the dressed nucleon.}
\label{polver}
\end{figure}

A general recipe for
calculating the in-medium coupling of the external potential to dressed
nucleons is known \cite{kadBaym}. 
The in-medium  vertex describing the coupling of an external perturbation to
 nucleons is given by the solution of the BS equation
(Fig. \ref{bsfig})
\begin{equation}
\Gamma=\Gamma_0+Tr K G_{ph} \Gamma \ ,
\end{equation}
 where $K$ denotes the particle-hole irreducible
kernel. The kernel $K$  of the BS equation 
should be taken consistently with the  chosen expression for the
self-energy, it is given by  the  functional derivative of the
self-energy with respect to the dressed Green's function 
 $K={\delta \Sigma}/{\delta G}$ \cite{kadBaym}.
Using the dressed vertex obtained as a solution of the BS equation the 
response function in the correlated medium can be obtained from the
diagram in Fig. \ref{polver}
\begin{equation}
\Pi_{(ST)}=Tr \Gamma^0_{(ST)} G_{ph} \Gamma_{(ST)} \ ,
\end{equation}
 $\Gamma_{(ST)}$ is the in-medium vertex for 
 the coupling in a given spin-isospin
$(ST)$ channel.
 The exact form of the BS equation in the real-time formalism (for
different types of vertices $\Gamma_{(ST)}$) can be found in 
\cite{pbpl,Bozek:2004ct}. The solution of the  BS integral equation is
obtained by iteration. It is a serious numerical task involving the
 calculation of two-loop diagrams with dressed propagators and broken
 rotational symmetry due to the presence of an external field.

The results for the response functions in different spin-isospin channels 
\cite{Bozek:2004ct} show that the calculation using dressed
propagators and dressed vertices obtained as solutions of the BS
equation is close to the RPA approximation. It is expected due to 
cancellations of self-energy and vertex corrections. Such cancellations occur 
for the exact solution of the system as well as within consistent
approximation derived from a  generating functional $\Phi$.
For the scalar residual interaction  the $\omega$-sum rule 
takes the
simple
form
\begin{equation}
\label{sumreq}
-\int \frac{\omega d \omega}{2\pi}{\rm Im}\Pi^r_{(ST)}({\bf q},\omega)=
\rho \frac{{\bf q}^2}{2m} \ ,
\end{equation}
in all the spin isospin channels ${(ST)}$. The above sum-rule severely
constraints acceptable forms of the response functions in the case of a scalar
residual interaction and leads to
very small multipair contributions in most of the kinematical regions.
Some differences can occur only when collective modes are present, 
as discussed latter.

The situation is very different for a more general form of the residual
interaction. 
\begin{figure}[ht]
\centering
\includegraphics*[width=0.6\textwidth]{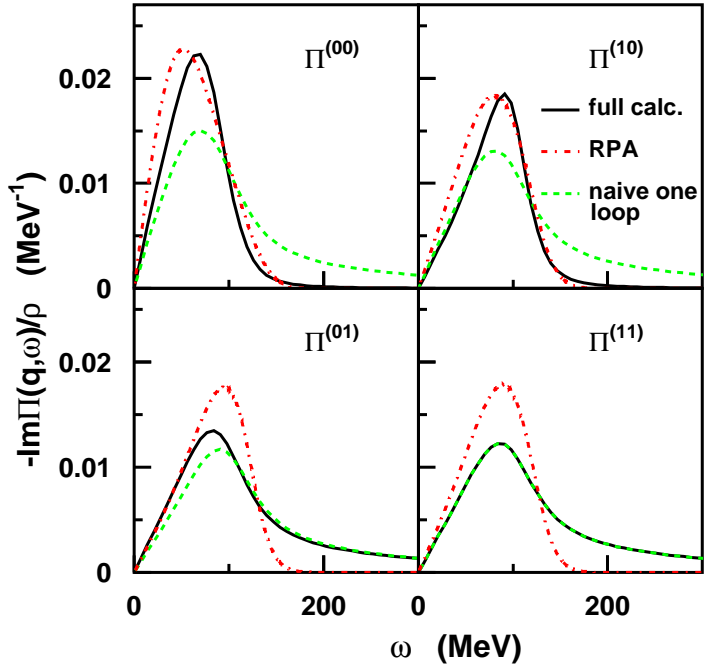}
\caption{The imaginary part of the polarization in the RPA approximation
  (dashed-dotted line), from the self-consistent calculation with 
dressed nucleons and vertices (solid line) and the naive one-loop polarization
  with dressed nucleons (dashed line). 
All results  are for an isospin dependent interaction $\frac{1}{2}(1+\tau_1\tau_2)V$,
  $q=210$MeV, and $T=15$MeV.}
\label{polall}
\end{figure}
 The kernel of the BS
equation depends very much on the isospin channel,
e.g. in  the channel $ST=11$ there are no vertex
corrections from the residual interaction at all. 
The propagators
are dressed by a nontrivial self-energy due to the residual interaction but
the vertex corrections are small or even absent. Therefore, the cancellation
between self-energy and vertex corrections, observed for a scalar interaction,
 can no longer be maintained. This has its implications for the $\omega-$sum
rule which gets modified for responses with $T=1$ \cite{Bozek:2004ct}.
In the case of the isospin dependent 
residual interaction,   in-medium dressed  nucleons
couple in the same way as  free nucleons to  isovector potentials.
In Fig. \ref{polall} the  response functions obtained with the isospin
dependent interaction are compared to
the response functions from the Fermi liquid theory.
For $T=0$ channels the correlated response function is similar to the
one particle-one hole response function. On the other hand,
the isovector response is  closer to the naive
one-loop result, without vertex corrections.

\begin{figure}[ht]
\centering
\includegraphics*[width=0.42\textwidth]{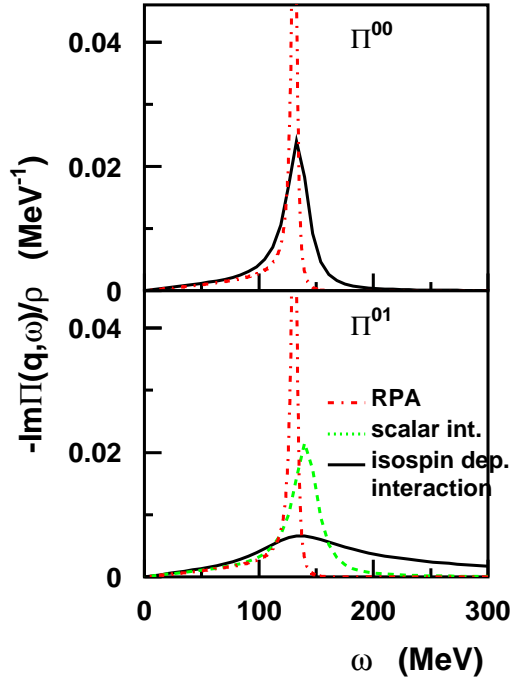}
\caption{The imaginary part of the polarization when a collective mode is
 present; density response (upper panel) and isospin response (lower panel).
The results are obtained in the RPA approximation (dashed-dotted line), and in
 the full calculation with dressed vertices for an isospin dependent residual
 interaction (solid line) and a scalar one (dashed line). For the density
 response the result is independent of the type of the residual interaction.}
\label{coll}
\end{figure}

To study the role of  multipair configurations on the collective
modes we increase the value of the Landau parameters.
Within
 the Fermi liquid theory
 a collective excitation at zero temperature is a discrete peak in the
imaginary part of the  response function. The state corresponding to
a collective excitation 
 cannot couple to incoherent one particle-one
 hole excitations. At finite temperature such a coupling is possible,
it can be calculated and the
 finite
temperature width of the collective state 
 is usually small. The collective state can acquire a finite width
 (also at zero temperature) due to the coupling to multipair
configurations \cite{pines}. The description of the damping of
collective states  from such processes  goes beyond the Fermi liquid theory. 

The result of the calculation with self-energy and vertex corrections is
compared to the RPA result in Fig. \ref{coll}. The collective mode in the
density response has a large width in the full calculation.
This corresponds to the coupling of the sound mode to multipair excitations
giving rise to a finite decay width, even at zero temperature (the small width
of the RPA collective mode is due to the finite temperature $T=15$MeV).
For the isovector response shown in the lower panel of Fig. \ref{coll} 
 the Fermi liquid theory predicts the presence of a well 
defined collective state (dashed-dotted line).
For a scalar residual interaction the response function in the
correlated system shows a collective state at similar energy (dashed line).
It has a larger width due to the contribution of multipair
configurations, analogously as in the density response.
On the  other hand, an isospin dependent residual interaction leads to
 an isovector  response with a large imaginary part at high energies 
(Fig. \ref{polall}).  For the chosen interaction we observe the extreme
 scenario of the disappearance of the collective mode when the effects of the
 residual interaction are taken into account (solid line in the lower panel of
Fig. \ref{coll}).
Generally we  expect a whole range
of behavior depending on the energy of the collective state and on
the strength of isospin dependent terms in the residual interaction but
always
the collective state is broader than in the Fermi
liquid theory, due to the coupling to multipair configurations. 
The same phenomena are expected also for the spin wave 
collective
state in the presence of spin dependent residual interactions.

\section{Summary}

 We present a new method of calculations for  nuclear matter.
The method discussed in this work is using in-medium Green's functions'
formalism and is based on the summation of diagrams with
a retarded propagation of a pair of fermions.
The sum of such ladder diagrams is the in-medium scattering matrix (the
 $T$-matrix). 
Self-consistency in this approach means that the fermion propagators in the 
$T$-matrix diagrams are dressed in a nontrivial self-energy. The self-energy
 itself is obtained from the sum of ladder diagrams.
This self-consistency requires the use of full spectral functions
depending on the momentum and energy of the particle.
The propagation off the mass shell, i.e.
the full dependence of the nucleon propagator on the energy, 
is a serious difficulty in numerical applications.
The progress achieved in the last years allows to perform extensive
calculation of the nuclear matter properties in the $T-$matrix approach at
zero and finite temperatures.

Attractive nuclear interactions lead to the formation of a superfluid phase at
low temperatures. Cooper pairs of fermions are formed, analogously as
for the electron superconductivity in metals.
The second important achievement here reported is the generalization of the
ladder
diagram summation to the superfluid phase. We discuss equations which allow
for a simultaneous and consistent treatment of the
short range nuclear interactions, the bound state formation (Cooper pairs) 
in the $T$-matrix and the superfluidity in nuclear matter.
Model calculations have allowed us to estimate the influence of the
superfluidity on standard  many-body effects in nuclear matter.
At the same time, the influence of many-body corrections on the 
superfluid gap can be assessed. These corrections, reducing the value of the
gap,
 can be described 
by an effective superfluid gap equation, similar to the mean-field 
Bardeen-Cooper-Schrieffer 
equation with an effective interaction and with a renormalized
value for the order parameter. 

The self-consistent $T$-matrix approximation is a thermodynamically consistent
approximation,
a so called conserving approximation. 
Single-particle properties obtained within this approximation are consistent
with global quantities describing the system, such as the binding energy or
the pressure.  Unlike  other methods, it allows to obtain directly
single-particle properties~:
 optical potentials, single-particle widths,
and spectral functions. These observables can be compared to experimental
values.

The calculation of processes occurring in the dense medium, e.g. particle
absorption, emission, neutrino cross-sections, requires the
knowledge of different kinds of linear responses of the system to  
external probes. In a correlated system, described using dressed propagators,
 the
calculation of the linear response is a very difficult task. The problem lies
in the need for a consistent dressing of the vertices corresponding to the
coupling of dressed nucleons to an external field. The in-medium vertices are
obtained from a solution of the BS equation, where the kernel of the equation
is derived from the same generating functional $\Phi$ as the self-energy.
 This procedure guarantees the fulfillment of conservation laws and sum-rules.
It must be contrasted with incomplete procedures with only self-energy effects
without vertex corrections. The linear response with self-energy and vertex
corrections beyond the Hartree-Fock+RPA approximation includes 
the effects of multipair excitations. For fermions interacting with a scalar
potential such effects are small, except for the damping of collective modes,
where the multipair configurations are important. The situation is very
different for spin or isospin dependent interactions. In that case the spin or
isospin response is very different form the RPA result; 
multipair effects are essential.

Applications discussed in this talk are restricted to nuclear systems. We
note that the same methods can be applied to other many-body fermionic
systems, high $T_c$ superconductors, fermions near  the Feshbach resonance,
the electron gas.

\vskip .3cm
This work was partly supported by the Polish State Committee for Scientific
Research Grant No. 2P03B05925.

\bibliography{../mojbib}

\begin{thebibliography}{10}
\expandafter\ifx\csname url\endcsname\relax
  \def\url#1{\texttt{#1}}\fi
\expandafter\ifx\csname urlprefix\endcsname\relax\def\urlprefix{URL }\fi

\bibitem{matsubara}
T.~Matsubara, Prog. Theor. Phys., {\bf 14} (1955) 351.

\bibitem{keldysh}
L.~V. Keldysh, Zh. Eksp. Teor. Fiz., {\bf 47} (1964) 1515.

\bibitem{Baym}
G.~Baym, Phys. Rev., {\bf 127} (1962) 1392.

\bibitem{KadanoffBaym}
L.~Kadanoff, G.~Baym, {\em Quantum Statistical Mechanics}, Bejamin, New York,
  1962.

\bibitem{di1}
W.~H. Dickhoff, Phys. Rev., {\bf C58} (1998) 2807.

\bibitem{Bozek:1998su}
P.~Bo\.zek, Phys. Rev., {\bf C59} (1999) 2619.

\bibitem{Dickhoff:1999yi}
W.~H. Dickhoff, C.~C. Gearhart, E.~P. Roth, A.~Polls, A.~Ramos, Phys. Rev.,
  {\bf C60} (1999) 064319.

\bibitem{Dewulf:2000jg}
Y.~Dewulf, D.~Van~Neck, M.~Waroquier, Phys. Lett., {\bf B510} (2001) 89.

\bibitem{Dewulf:2002gi}
Y.~Dewulf, D.~Van~Neck, M.~Waroquier, Phys. Rev., {\bf C65} (2002) 054316.

\bibitem{Bozek:2002em}
P.~Bo\.zek, Phys. Rev., {\bf C65} (2002) 054306.

\bibitem{Bozek:2002tz}
P.~Bo\.zek, P.~Czerski, Acta Phys. Polon., {\bf B34} (2003) 2759--2768.

\bibitem{Frick:2003sd}
T.~Frick, H.~{M\"uther}, Phys. Rev., {\bf C68} (2003) 034310.

\bibitem{Alm:1996ps}
T.~Alm, G.~{R\"opke}, A.~Schnell, N.~H. Kwong, H.~S. Kohler, Phys. Rev., {\bf
  C53} (1996) 2181.

\bibitem{Bozek:2001tz}
P.~Bo\.zek, P.~Czerski, Eur. Phys. J., {\bf A11} (2001) 271.

\bibitem{Bozek:2002mw}
P.~Bo\.zek, Eur. Phys. J., {\bf A15} (2002) 325.

\bibitem{Dewulf:2003nj}
Y.~Dewulf, W.~H. Dickhoff, D.~Van~Neck, E.~R. Stoddard, M.~Waroquier, Phys.
  Rev. Lett., {\bf 90} (2003) 152501.

\bibitem{3holes}
H.~Q. Song, M.~Baldo, G.~Giansiracusa, U.~Lombardo, Phys. Rev. Lett., {\bf 81}
  (1998) 1584.

\bibitem{Bozek:2003wh}
P.~Bo\.zek, Phys. Lett., {\bf B586} (2004) 239.

\bibitem{Frick:2004th}
T.~Frick, H.~{M\"uther}, A.~Rios, A.~Polls, A.~Ramos, Phys. Rev., {\bf C71}
  (2005) 014313.

\bibitem{Bozek:2002ry}
P.~Bo\.zek, P.~Czerski, Phys. Rev., {\bf C66} (2002) 027301.

\bibitem{BCS}
J.~Bardeen, L.~N. Cooper, J.~R. Schrieffer, Phys. Rev., {\bf 108} (1957) 1175.

\bibitem{Vonderfecht:1991}
B.~Vonderfecht, C.~Gerhart, W.~Dickhoff, A.~Polls, A.~Ramos, Phys. Lett., {\bf
  B253} (1991) 1.

\bibitem{Bozek:2001nx}
P.~Bo\.zek, Phys. Rev., {\bf C65} (2002) 034327.

\bibitem{Bozek:1999rv}
P.~Bo\.zek, Nucl. Phys., {\bf A657} (1999) 187.

\bibitem{schrieffer}
J.~R. Schrieffer, {\em Theory of superconductivity}, W. A. Benjamin, New York,
  1964.

\bibitem{Bozek:2000fn}
P.~Bo\.zek, Phys. Rev., {\bf C62} (2000) 054316.

\bibitem{Bozek:2002jw}
P.~Bo\.zek, Phys. Lett., {\bf B551} (2002) 93.

\bibitem{Haussman2}
R.~Haussmann, Z. Phys., {\bf B91} (1993) 291.

\bibitem{Knoll:1996nz}
J.~Knoll, D.~N. Voskresensky, Annals Phys., {\bf 249} (1996) 532.

\bibitem{pines}
D.~Pines, P.~Nozi\`eres, {\em The Theory of Quantum Liquids Vol. I}, Benjamin,
  New York, 1966.

\bibitem{olpe}
E.~Olsson, C.~J. Pethick, Phys. Rev., {\bf C66} (2002) 065803.

\bibitem{gogny}
D.~Gogny, R.~Padjen, Nucl. Phys., {\bf A293} (1977) 365.

\bibitem{bonitz}
N.~Kwong, M.~Bonitz, Phys. Rev. Lett., {\bf 84} (2000) 1768.

\bibitem{faleev}
S.~Faleev, M.~Stockman, Phys. Rev., {\bf B66} (2002) 085318.

\bibitem{pbpl}
P.~Bo\.zek, Phys. Lett., {\bf B579} (2004) 309.

\bibitem{henn}
D.~Tamme, R.~Schepe, K.~Henneberger, Phys. Rev. Lett., {\bf 83} (1999) 241.

\bibitem{kadBaym}
G.~Baym, L.~Kadanoff, Phys. Rev., {\bf 124} (1961) 287.

\bibitem{Bozek:2004ct}
P.~Bo\.zek, J.~Margueron, H.~{M\"uther}, Ann. Phys., {\bf 318} (2005) 245.

\end{thebibliography}

\end{document}